# QoS Provision for Controlling Energy Consumption in Ad-hoc Wireless Sensor Networks

Ahmed A. Mawgoud [1*], Mohamed Hamed N. Taha [1], Nour Eldeen M. Khalifa [1]

[1] Information Technology Department, Faculty of Computers and Artificial Intelligence, Cairo University, Giza, Egypt

[*] <u>aabdelmawgoud@pg.cu.edu.eg</u>

**Abstract:**
Ad-hoc wireless sensor network is an architecture of connected nodes; each node has its main elements such as sensors, computation and communications capabilities. Ad-hoc WSNs restrained energy sources result in a shorter lifetime of the sensor network and inefficient topology. In this paper, a new approach for saving and energy controlling is introduced using quality of service. The main reason is to reduce the node's energy through discovering the best optimum route that meets QoS requirements; QoS technique is used to find the optimum methodology for nodes packets transmission and energy consumption. The primary goals of the research are to discover the best techniques to 1) Minimize the total consumed energy in the ad hoc wireless sensor network. 2) Maximize the ad hoc wireless sensor network lifetime. The simulations of the problem will be formulated with the use of Integer Linear Programming.

## 1. Introduction

The Wireless Ad-hoc wireless sensor network is categorized as a decentralized network; It is Ad-hoc due to the fact it is having a lack of a base station; as there is no a pre-architecture infrastructure to depend on [1]. In the Ad-hoc multi-hop networks, any connection between two nodes that no longer exist inside transmission range, the packets need to be relayed via using the route of the intermediate nodes. Each node acts as a host and a router as well [2]. (Poovendran and Lazos, 2006) stated in their study that every node is in the routing technique with passing the facts from one to other nodes and so the nodes can send the created in a dynamic way depending on the network connectivity [3]. There are four fundamental elements in a sensor network [4], as shown in figure 1:

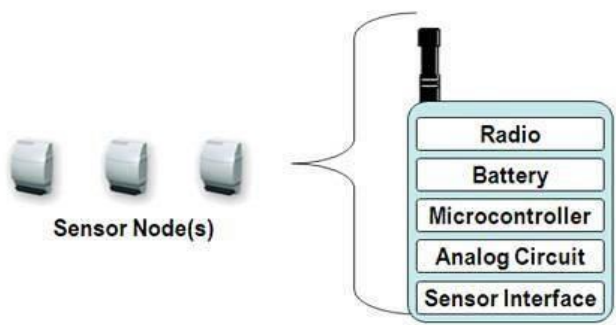

*Fig. 1.  Ad-hoc WSN sensor node components*

### 1.1  Ad hoc wireless sensor network components

- **Radio:** It is a wireless cognitive allotted network of radio sensor nodes that has the ability to sense an event signal and collaboratively communicate their readings over dynamically to be had spectrum bands in a multi-hop way to meet the application-particular necessities [5].
- **Battery:** They are battery-powered devices; given that it is commonly tough or not possible to run main resources to their deployment site. Energy to the wireless sensor nodes is normally provided via primary batteries [6].
- **Microcontroller:** It performs duties, processes records, and controls other components' capability in the sensor node [7].
- **Analog Circuit:** It represents an essential part in sensor sign acquisition because of the analog nature of sensory indicators [8].
- **Sensor Interface:** They are hardware devices that mainly used to capture records from the environment and produce a measurable change response in physical conditions [9].

The battery is the second component used in the ad hoc WSNs; the trade-offs should be a main factor; because of the high traffic rate, a high amount of energy is being consumed. we have to put into consideration the long-life requirements of according to the size and the weight besides the common criteria for battery availability [10]. In order to extend the battery lifetime, a node of the wireless sensor network should transmits the packets iteratively after powering on the radio. After transmitting, it powers the radio off to preserve energy [11]. Ad hoc WSN radio transmits the signal efficiently and return the lowest energy use. As a result, the processor involved has the capability to manage extra effects such as waking, powering up and returning to sleep mode. The microprocessor in ad hoc WSNs consists of decreasing energy consumption while keeping or increasing processor speed [12]. Figure 2 represents an ad hoc WSN common hierarchy.

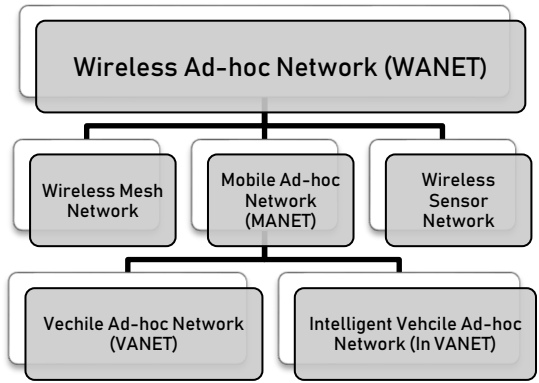

*Fig. 2.* *Wireless ad-hoc networks common topologies*

The wireless ad hoc sensor network technology is used widely in various industries through developed applications. These technologies grew to become accessible as the sensor became smaller, sophisticated and cheaper [13]. Those sensors were designed with wireless interfaces; in order to have the ability of communication from node to another to shape a full network [14]. Many elements are affecting substantially the design of the Ad-hoc WSN such as the environment, the hardware components, the software design, the availability and the constraints [15]. (X. Cheng et al, 2003) defined in their study that the ad hoc WSN is consisting of many independent nodes that are connected with each other to form a wireless ad hoc network. Sensors are being used in this network for monitoring physical conditions. The Ad-hoc WSN contains a gateway that connects to the wired world and different nodes by using wireless connectivity as shown in figure 3. Each application's requirement needs a well-chosen wireless protocol [16].

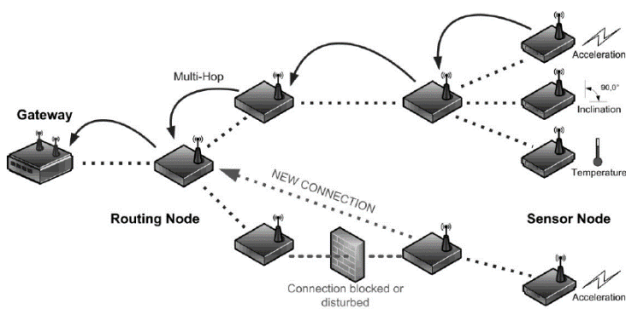

*Fig. 3.* *An exapmle of ad hoc wireless sensor network architecture (Fernandez-Steeger et al., 2009)*

Ad-hoc WSN technology offers a lower value system because of lowering energy and providing a higher resource management approach [17]. Remote monitoring can minimize the network costs; due to the fact, it can connect wired networks with wireless ones to open the way for measurement applications, there are three types of network topologies. They are used in the design of any ad hoc WSN nodes, as shown in figure 4.

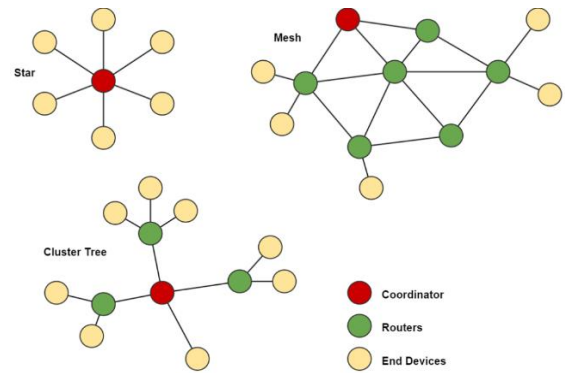

*Fig. 4.* *Wireless ad-hoc networks common topologies*

### 1.2. Types of network topologies

1.2.1 *Star Topology*: it is the simplest structure of the network topologies; because every node in this network connects to a central device (e.g. Computer, Switch or Router). The nodes act as clients. Star topology is very common; due to the fact, it is easy to add another device to the network. If one node fails, the rest of the network continues working normally.
1.2.2 *Cluster Tree Network:* In this network, each node connects to a higher node in a tree ending with a gateway. The route of the records begins from the lowest node ending with the gateway.
1.2.3 *Mesh Topology:* (Peterson and Davie, 1998) concluded in their study that in order to provide a reliability increment, using mesh networks would provide the node -that has the capability to connect to multiple nodes at the identical time- the best routing path; that make the topology [18].

## 2. Related Work

Many related works have studied the topology control for wireless ad hoc sensor networks. Firstly, the relationship between the throughput and transmission range in topology control was studied; to permit transmitting energy adjustment to decrease interference, there is a need for developing an analytical model to obtain high throughput. (A Tiwari et al, 2007) stated in their experiment that there is no focal point in each work for minimizing the consumed energy. There had been earlier topology control works that aim to decrease nodes' interference and attaining high throughput via adjusting every node's transmitting electricity an analytic model [19]. Table 1 below is a comprehensive study about previous studies with various techniques to consume energy in wireless sensor networks in general.

**Table 1** Study on energy consumption challenges in WSNs

| Ref No. | Main Challenge | Technique | Type |
|---|---|---|---|
| A Manjeshwar et al. (2001) [20] | Enhancing energy efficiency in Wireless Sensor Networks | TEEN Routing Protocol | Technical |
| Y Yu et al. (2001) [21] | Forwarding a packet to nodes in a geographic region | GEAR Algorithm | Technical |
| S Cui et al. (2004) [22] | Analyse best transmission strategy for minimizing energy | MIMO Technique | Technical |
| I. Kadayif et al. (2004) [23] | Strike a balance in computation & communication energy | Sensor Data Filtering | Technical |
| I Gupta et al. (2005) [24] | Cluster-head election in wireless sensor networks | Fuzzy Logic | Technical |
| M Gabriela et al. (2006) [25] | Mica2 motes using GSP for routing and a CSMA | Gossip-Based Sleep | Technical |
| F Shu et al. (2006) [26] | Consuming energy in the medium access control (MAC) | Developed Framework | Technical |
| F Shebli et al. (2007) [27] | Reduce consumed energy within linear sensor networks | New Proposed Algorithm | Technical |
| G Küçük et al. (2007) [28] | Reduce processor component energy dissipation in WSNs | MC/CADMA Architecture | Technical |
| T Wang et al. (2008) [29] | Minimize energy consumption per information bit link, packet retransmission & overhead. | AWGN Channel | Theoretical |
| Halgamuge et al. (2009) [30] | Comprehensive energy model for WSNs | Modified LEACH Model | Technical |
| Y Liang et al. (2010) [31] | Explore the temporal correlation in monitoring WSNs | Two-Modal Transmission | Technical |
| Y Wang et al. (2010) [32] | Stochastic analysis of energy consumption in WSNs | Stochastic Queue Model | Theoretical |
| JM Molina et al. (2010) [33] | Energy estimator and profiler in WSNs | Transaction Level Model | Technical |
| X Wang et al. (2011) [34] | Coverage investigation of mobile heterogeneous WSNs | Equivalent Sensing Radius | Theoretical |
| K Oikonomou et al. (2012) [35] | Prolong wireless sensor networks lifetime | Fixed Sink Assignment | Technical |
| JS Guang et al. (2012) [36] | Ensure the existence of active sensors for service request | Sleeping Scheduling Model | Technical |
| T Ducrocq et al. (2013) [37] | Maximize the time with nodes to satisfy application needs | Battery Aware Clustering | Technical |
| NH Abbas et al. (2013) [38] | Consume energy in WSNs using MPSO & ACO | Nature-Inspired Algorithms | Technical |
| J Saraswat et al. (2013) [39] | Selecting nodes with high energy close to the base station | New Proposed Algorithm | Technical |
| LV Andreyevich et al. (2013) [40] | The ability to save energy in wireless sensor networks | LTE with MIMO | Technical |
| H Cotuk et al. (2014) [41] | Investigate the limited bandwidth impact on WSN energy | LP Model | Technical |
| M Nandi et al. (2014) [42] | Multi-layer medium access control (ML-MAC) scheme | Swarm Optimization | Technical |
| L. Catarinucci et al. (2014) [43] | Implement a duty-cycle-based communication protocol | Cross-Layer Approach | Technical |
| G Awatef et al. (2014) [44] | Maximize the lifetime of wireless sensor networks | MEDD-BS Clustering | Technical |
| Rui Hou et al. (2014) [45] | Calculate the battery consumed energy in WSNs | Path Correlation | Technical |
| SZ Yazdanpanah et al. (2014) [46] | Increase the WSNs energy & longevity | New Proposed Algorithm | Theoretical |
| A Girgiri et al. (2015) [47] | Consume energy using a sensor's transmitter radio model | LEACH Protocol | Theoretical |
| M Lounis et al. (2015) [48] | Simulation times faster than sequential models | GPU Architecture | Technical |
| S Yessad et al. (2015) [49] | Balancing the energy consumption for each node | Cross-Layer Routing | Theoretical |
| B Amutha et al. (2015) [50] | Latency reduction through load balancing | Eco-sense Protocol | Technical |
| Seyyed et al. (2016) [51] | Increase the wireless sensor networks' lifetime | CICA Algorithm | Technical |
| V Toldov et al. (2016) [52] | The relation between interference & energy consumption | X-MAC RDC Protocol | Technical |
| H Oudani et al. (2017) [53] | Consume energy when transmit data in the Base Station | Clustering LEACH Protocol | Technical |
| M Okopa et al. (2017) [54] | Analytical Model of delay & average energy consumption | M/G/1 Queue Model | Technical |
| H Alhumud et al. (2018) [55] | reduce the energy consumption and increase the network lifetime in multiple numbers of greenhouses | CRSN sensing protocol | Technical |
| R Alhussaini et al. (2018) [56] | Reduce the redundant data during aggregation in WSNs | Data Transmission Protocol | Theoretical |
| MC Blajer et al. (2018) [57] | Maximal battery life-time in WSNs | Stochastic Algorithm | Theoretical |
| KS. Ananda Kumar (2018) [58] | Energy consumption and packet delivery ratio in WSNs | Itree-MAC Protocol | Technical |

## 3. Problem Statement

Any ad hoc WSN consists of a group of transceivers. The entire communications between the transceivers are primarily based on the radio broadcast. There is existence for the transmission energy. Transmission relies on many factors such as the transceiver distance, the antenna's sender route, collision, etc.

(Tabaa, 2016) stated that the nodes are corresponding "one to one" in transceivers. The (*x,y*) points exist directly in the design if only the transmission strength (*x*) has the lowest sending energy threshold. The main topology control goal is to assign the transceivers' needed energy. As a result, it is expected that the output design would match the targeted properties. Although, every transceiver's resource battery energy is expensive, it is suggested that to reach the needed goal throughout minimizing the transmission energy given function [59]. (Chen, Makki & Pissinou, 2009) their research objective was to study the transceivers' main energy and the entire transceivers' final energy (the remaining objective targets to limit the assigned transceiver average). To this point, the main motivation behind analyzing the problems of topology control is to reach the best usage from each node's energy. Additionally, using every node's minimum energy to achieve a given task will minimize the MAC to the closest signals [60].

### 3.1. Formulation of Topology Control Problems

(Aggelou, 2004) stated that the primary topology control purpose is to identify the energy of transmission to nodes; as a result, the output of the undirected design has a unique property and a representative feature to the minimized energy which used to be assigned to the node. In his study, the problem of topology control was classified into two models of graphs:
- Direct Graph Model.
- In-direct Graph Model.

The (Directed Graph Model) ease the communication in two-ways between the pairs and the communication in one-way between the other pairs. On the contrary, each edge in the (In-direct Graph Model) is corresponding to communication in many ways [61].

## 4. Solution Methodology

The Ad-hoc Wireless Sensor Networks are naturally decentralized which make them suitable for various types of applications where there is no relying on central nodes, which means that the scalability of the ad-hoc networks will be greater than the ordinary wireless managed the network. For ad-hoc WSNs, the limits of theoretical and sensible for the typical capacity had been identified. The Ad-hoc WSN is formed from many connected nodes via links. Links depend on the node's resources (e.g. memory, transmission energy, and computing energy), the behavioral properties and the connection properties [62]. Tactlessly, there is a possibility for a dis-connection to happen in the network at any time, the operational network needs to be having the capability to overcome these dynamic risks in an efficient, scalable and secure method. Mostly, the nodes in ad-hoc WSNs are competing for accessing the shared medium. Because of this competition, collision interference will occur. To resolve this problem there be an enhancement for the immunity the usage of the co-operative wireless communications. It solves the interference through having the destination node combines self-interference and other nodes interference in order to enhance the preferred signal decoding [63]. Nowadays, due to the rapid-growth development in fields like IoT, big data, deep learning there is a need for new schemes to provide energy consumption; QoS is one of the solutions for consuming energy in ad hoc WSNs (e.g. Multimedia Transmission, Real-Time Work and Interactive Applications) and effective network topology. The major QoS goal is to make the network conduct deterministically, as a result, the records delivering via this network in a better way and making use of the resources in a better way [64]. To make sure that the most essential QoS parameters had been met, the end-to-end loop should be provided using the QoS routing method with free paths.

### 4.1. Parameters

#### 4.1.1. QoS Parameters

1) Battery Existence-Time.
2) Bandwidth.
3) Network Availability.
4) Communication Groups.
5) Contingency Services.

#### 4.1.2. Network Parameters

1) The Battery.
2) The Delay Jitter.
3) Buffering Space.
4) Process Strength.
5) The Bandwidth.

### 4.2. Routing Protocols Classification (QoS)

Because of the low range of every transmitter effectiveness, multi-hop paths are used through distant nodes to communicate with other nodes. As a result, obtaining a high availability performance in an environment where the change of the network occurs dynamically, there is a declaration on giving the shortest route between every two nodes. However, all the solutions deal only with traffic.

Eventually, a given network graph *O* and the traffic demands between node pairs, the traffic needs to be routed away that achieves the purpose of the experiment by reducing the consumed energy in the whole network. In the system, the maximum load node that is referred to by using L*max, x* needs to be minimized. The components of this problem are represented by using the set of equations as presented in the formula below (1) - (5) [65]:

$$Min\ L_{max} \qquad (1)$$

$$L_{max} \geq 0 \qquad (2)$$

$$\sum_i f_{i,j}^{s,d} - \sum_i f_{j,i}^{s,d} = \begin{cases} \lambda_{s,d} & if\ s = i \\ \lambda_{s,d} & if\ d = i \\ 0 & otherwise \end{cases} \qquad (3)$$

$$\sum_{(s,d)} \sum_j f_{i,j}^{s,d} + \sum_{(s,d)} \sum_j f_{j,i}^{s,d} \le L_{max} \quad (4)$$

$$f_{i,j}^{s,d} \ge 0, \quad \forall\, i,j \in V, (s,d) \quad (5)$$

$$Note\ that: \forall (s,d), f_{i,j}^{s,d} = 0$$

From the formulation above, for every $f_{j,i}^{s,d} = 0$, Min L*max* represents the objective that is responsible for minimizing the maximum node value for bandwidth utilization. Function (4) sets the factor (i.e., L*max*) of bandwidth is used by node i. solving the whole formulation (1) - (5) will obtain the value of L*max*. Capacity exceeding can be identified for some nodes bandwidth usage when the value of L*max* > 1

### 4.3. Solution Design

#### 4.3.1. Given

- P, maximum allowed transmission power.
- B, node bandwidth.
- $\lambda_s$, the node (s, d) traffic demand.
- V, a set of nodes with each node's location.
- $\Delta_s$, the maximum hop count.

#### 4.3.2. Variables

- $x_{i,j}$, (i,j) are Boolean,
- $x_i = 1$ if there is a link from node i to node j; otherwise, $x_i = 0$.
- s, Boolean Variables, $xs, =1$; otherwise $xs, d= 0$.
- E*max*, The node maximum transmitting energy.

#### 4.3.3. Constraints

- Topology Constraints

Function 6: Reduce the maximum transmission energy.

$$Min\ E_{max} \quad (6)$$

Function 7: Ensure the availability of two direct links for each edge.

$$x_{i,j} = x_{j,i} \quad \forall\, i.j \in v \quad (7)$$

Function 8: Ensure the broadcast ability for the nodes when the node transmits all the nodes in the transmission range will receive from this node.

$$x_{i,j} \le x_{i,j'} \quad If\ d_{i,j'} \le d_{i,j} \quad \forall\, i.j.j' \in v \quad (8)$$

- Transmitting Power Constraints

Function 9: Identify the maximum transmission energy for all nodes.

$$e \ge e_{max} \ge d_{i,j}^a\, x_{i,j} \quad \forall_i < j. i. j \in v \quad (9)$$

Function 10: Ensure that there is no exceeding from the hop-count to the pre-specified bound.

$$\sum_{(i,j)} x_{i,j}^{s.d} \le \Delta_{s.d} \quad \forall\, (s.d) \quad (10)$$

- Bandwidth Constraints

Function 11: The total signals and transmission at a node do not exceed the bandwidth capacity of this node.

$$x_{i,j}^{s.d} \le x_{i,j} \quad \forall\, i.j \in V \quad (11)$$

Function 12: Ensure the route validity for each node-pair; as the traffics are not splittable $x_{i,j}$ represents the entire traffics of (s,d) going through the link (i, j) if it is in the route from s to d.

$$\sum_j x_{i,j}^{s.d} - \sum_j x_{j,i}^{s.d} = \begin{cases} 1 & if\ s = i \\ -1 & if\ d = i \\ 0 & otherwise \end{cases} \quad (12)$$

- Binary Route Constraint

$$x_{i,j} = 0,\ or\ 1, x_{i,j}^{s.d} = 0\ or\ 1 \quad (13)$$

#### 4.3.4. Threshold Constraint

The main target of our contribution is to add a new constraint that can affect the consumed energy of the transmitted data. Alternatively, the variance in the energy matrix should be reduced; the main reason of this constraint is using the minimum energy transmission for each node, by using the nodes that has not consumed its average energy.

$$E_i \le E_{average} + Threshold \quad \forall\, 1 \le i \le n \quad (14)$$

There were two problems during integrating this constraint into the simulation:
- The added constraint tighten the system and provide no strong benefit to the sensor network without an existence of ta threshold, to solve this problem a threshold value was added to allow having higher number of nodes above the average consumed energy.
- The average consumed energy in the first three requests are always equal to the nodes energy transmission, which fails to achieve the main goal of the constraint. However, after the third request the ratio of consumed energy appeared and showed the constraint effectiveness.

With a feasible delay route and the accurate state information, the routing algorithms were mainly designed to resolve these problems. The unique properties of the algorithms are explained as follows:

1. *High-level tolerance for the non-accurate available information:* An extreme performance is accomplished with the lack of correct state information, based on (Average Path Cost, Success Ratio).
2. *Increasing the probability of achieving the possible route by using a multi-path routing:* The search can be achieved for a limited number of paths that restrict the routing above. For each node, it is important to collect data to reach the optimal routing path.
3. *The Optimal Routing Path should be considered:* Enhancing the overall performance of the entire sensor network

The next section is the experimental section to prove the importance of the optimum threshold in the sensor network topology with a discussion about the effect different threshold values effect.

## 5. Experimental Section

The simulation design was made in a 180×180 two-dimensional region [31]. The number of assumed nodes were 15 nodes as shown in figure 5 below; every node in this topology is having its own range as well as its neighbours' nodes that can intersect with their range. The model was designed as an integer linear programming using Matlab, it is also ready to deal with any number of nodes and the only difference will be with the compilation time complexity using the previously mentioned equations (1-8).

1. The nodes coordinates were uniformly distributed inside the region.
2. The transmitting energy value function is set to 2.
3. For each node, a random Poisson function with the mean value $\lambda = 1$ to generate a number $k$, that represents the requests numbers from this node.
4. The destination requests are picked randomly via other nodes.

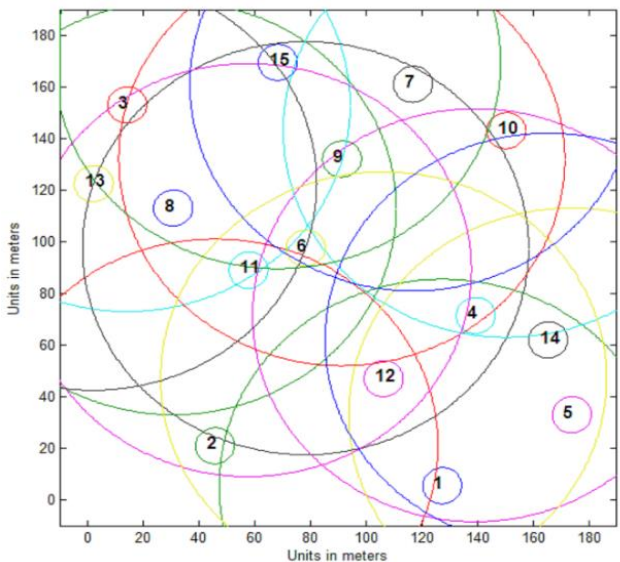

***Fig. 5.*** *Network topology of fifteen nodes in 180x180 m*

### 5.1. Topologies versus Traffic Load:

The optimum threshold was the first problem to test in this experiment, so there was a need to run the model simulation to assign the relation between the threshold -with different $\lambda m$ values- and the variance that was represented in both Figures (5, 6); to prove that the topology control problem is an optimization problem. Figure 6 describes the minimum variance at a threshold more than or equal to 100 for $\lambda m = 37.5$.

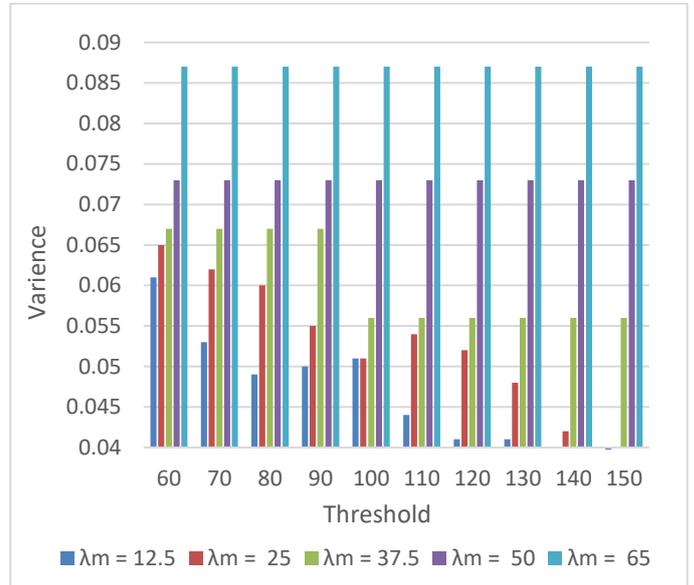

***Fig. 6.*** *Variance vs Threshold Values with different $\lambda m$ Values*

Figure 7 describes the relation between the varience and the λ values, it shows the instability of the sensor network occures when λ value become higher. This can be seen from the varience difference values when λ started with 60 and ended with 120.

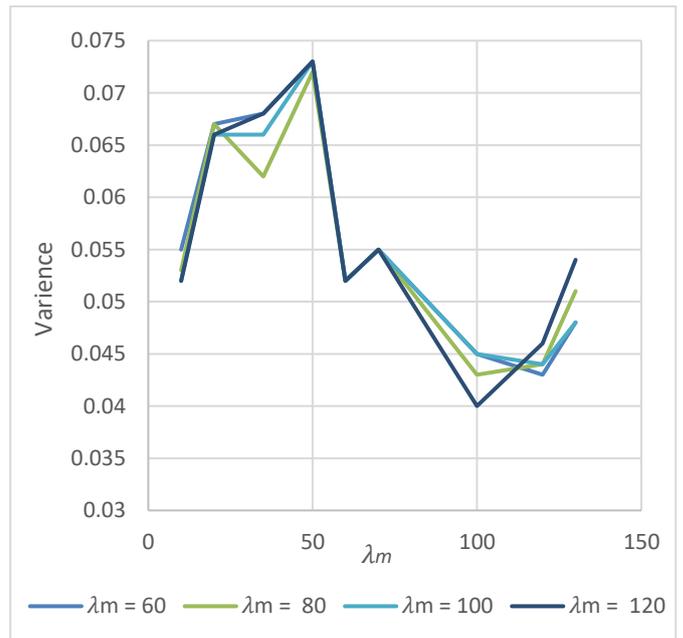

***Fig. 7.*** *Comparison between variance & $\lambda_m$ values*

As shown in figure 8 below, a relation between the lost packets number with different λ values, it shows that the higher λ value becomes, the higher chance to lose packets. When λ was lower than 90 the system was stable and all the packets was sent successfully.

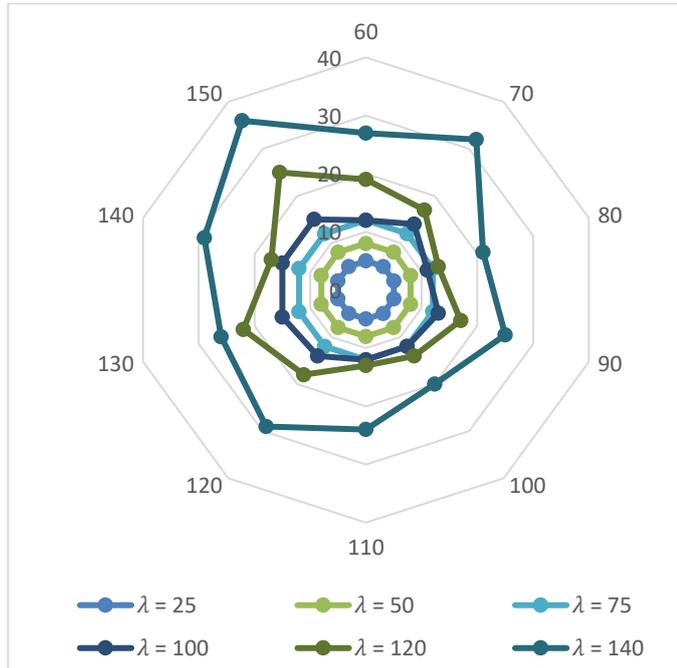

***Fig. 8.*** *Lost Packets vs $\lambda m$ values*

However, the system instability occurred when λ value becomes 100 or higher, the number of lost packets increased more and more with the λ values increment. From all the previous results, it is obvious that the problem of ad hoc WSNs sensor topology is an optimization problem, both threshold and λ were the factors measured to test the QoS effect on the network topology. The hop count indicates to the intermediate number of devices that the packets must go amongst source and destination; it is a hard to measure the two nodes distance. Tables (2, 3, and 4) represent 11 requests made in the simulation to define the optimum routing path between the sender and the receiver. In table 2, it describes the first routing compilation, all the packets were sent except seven packets for the request number (1, 5, 6, 7, 10 and 11).

**Table 2** $\lambda m = 10$, Threshold = 0, Hop count = 2, Variance = 0.034

| Req. # | $\lambda_{s,d}$ | Sender | Receiver | Routing Path |
|---|---|---|---|---|
| 1 | 1 | 1 | 9 | Lost |
| 2 | 3 | 3 | 8 | 3 → 8 |
| 3 | 4 | 4 | 1 | 4 → 1 |
| 4 | 4 | 4 | 12 | 4 → 12 |
| 5 | 4 | 4 | 7 | Lost |
| 6 | 5 | 5 | 8 | Lost |
| 7 | 5 | 5 | 15 | Lost |
| 8 | 10 | 10 | 5 | 11 → 9 |
| 9 | 8 | 11 | 9 | Lost |
| 10 | 9 | 12 | 3 | Lost |
| 11 | 10 | 12 | 13 | Lost |

However, the need to know the threshold constraint effect on the packet routing process raised the need to run the experiment again with higher constraint number. In table 3, when the threshold constraint value equals 3, all the packets were sent except three packets for the request numbers (6, 7,11)

**Table 3** $\lambda m = 9$, Threshold = 3, Hop count = 3, Variance = 0.0852

| Req. # | $\lambda_{s,d}$ | Sender | Receiver | Routing Path |
|---|---|---|---|---|
| 1 | 7 | 1 | 9 | 1 → 12 → 9 |
| 2 | 12 | 3 | 8 | 3 → 8 |
| 3 | 10 | 4 | 1 | 4 → 14 → 1 |
| 4 | 13 | 4 | 12 | 4 → 12 |
| 5 | 11 | 4 | 7 | 4 → 9 → 7 |
| 6 | 15 | 5 | 8 | Lost |
| 7 | 12 | 5 | 15 | Lost |
| 8 | 12 | 10 | 5 | 10 → 4 → 5 |
| 9 | 15 | 11 | 9 | 11 → 6 → 9 |
| 10 | 10 | 12 | 13 | 12 → 11 → 13 |
| 11 | 15 | 12 | 5 | Lost |

From table (2,3), the threshold numbers have a noticeable effect on the sent packets through the sensor network. It is concluded that the higher the constraint number becomes, the low possibility of losing packets in the network.

**Table 4** $\lambda m = 10$, Threshold = 7, Hop count = 3, Variance = 0.08524

| Req. # | $\lambda_{s,d}$ | Sender | Receiver | Routing Path |
|---|---|---|---|---|
| 1 | 9 | 1 | 9 | 1 → 4 → 9 |
| 2 | 14 | 3 | 8 | 3 → 8 |
| 3 | 13 | 4 | 1 | 4 → 12 → 1 |
| 4 | 11 | 4 | 12 | 4 → 12 |
| 5 | 15 | 4 | 7 | 4 → 10 → 7 |
| 6 | 13 | 5 | 8 | 5 → 4 → 8 |
| 7 | 14 | 5 | 15 | 5 → 10 → 15 |
| 8 | 14 | 10 | 5 | 10 → 4 → 5 |
| 9 | 12 | 11 | 9 | 11 → 6 → 9 |
| 10 | 11 | 12 | 13 | 12 → 6 → 13 |
| 11 | 12 | 12 | 5 | 12 → 14 → 5 |

## 6. Results and Discussion

The experiment proves that the consuming energy is one of many factors (i.e. software applications and hardware components) that affect the lifetime of the wireless sensor networks. we can conclude that:

1. When the threshold has a higher value, the more stable the sensor network topology becomes.
2. When λ has a lower value, the topology performance has a better impact and less lost packets rate.
3. The efficiency of the constraint increases with the nodes increases.

The results of the experiment are limited to the mentioned constraints that was already mentioned in the problem formulation –it is changeable based on the network scale and function-. However, the added threshold constraint showed effectiveness in node energy consumption during the execution for the ad hoc wireless sensor networks. As a result, the general idea of consuming energy for nodes in ad-hoc wireless sensor networks using QoS proved its efficiency in maximizing the overall lifetime of the ad-hoc WSN. A comparison was made between the proposed QoS approach and other approaches such as MIMO, Swarm Optimization and LEACH, that was proposed in [40][42][47] respectively to prove the efficiency in consuming energy and stability of sending all packet without high failing rate. Figure 9 represents the overall consumed energy between our approach and the proposed results in [40][42][47].

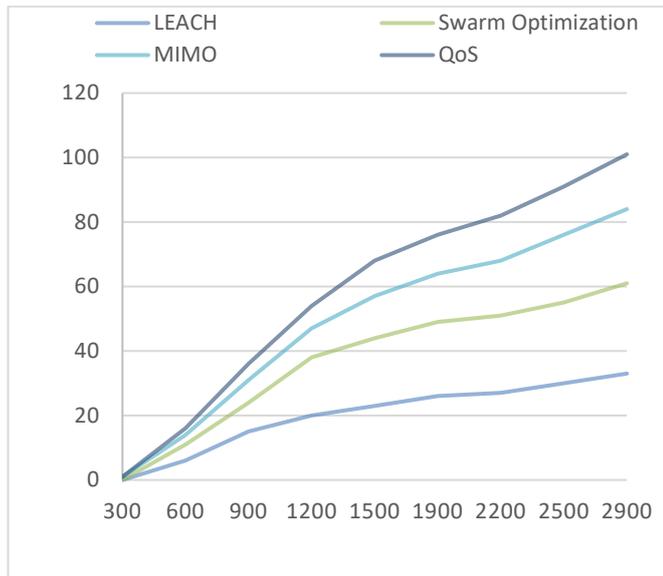

*Fig. 9.  A comparison of total consumed energy during the excecution in the ad hoc WSN*

As it is shown in the figure above, the proposed MIMO approach in [40] was the highest approaches consuming energy in all the rounds for ad hoc WSN systems with only 30 joule. However, there is a convergence of results between MIMO approach and QoS with XX Joule difference. Which proves that QoS provided the needed consuming energy with a door of improvement for new methodologies between QoS and MIMO approaches.

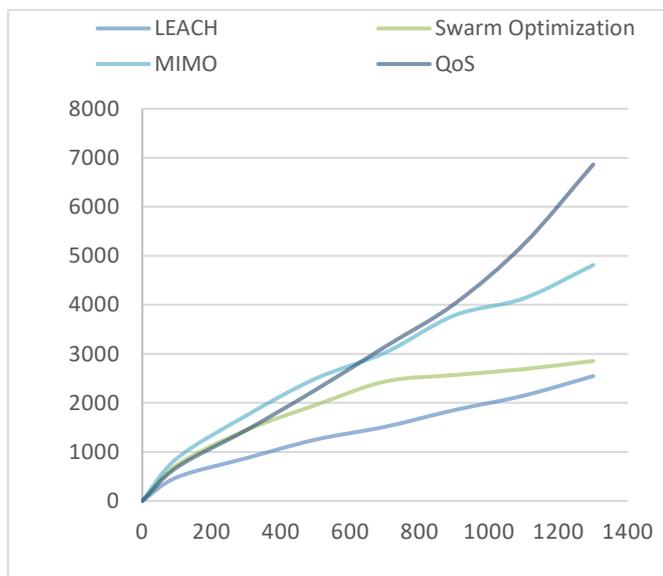

*Fig. 10.  A comparison of sent packets numbers in each round in the ad hoc WSN*

figure 10 compares between the successful sent packets for all four approaches in 1300 rounds. LEACH and Swarm Optimization methods were the lowest rate approach with XX and XX successful sent packet. Although, QoS provided a hgih rate of sucsesful sent packets with XX packets. It is stil arguable about the results in lrge-scale adhoc wireless sensor networks for all approaches after long time (e.g years and decades). Consuming the energy is one factor of maximizing the node lifetime in the ad-hoc wireless sensor networks. However, there are many other -software and hardware- factors that affects the overall network lifetime, a future work can be an open door for adding new constraints in QoS approach to cover most of the vulnarabilities in large-scale ad hoc WSN in long period of time.

## 7. Conclusion

This paper investigated the problem of topology control using the QoS to be able to reduce the consumed energy between the nodes in the wireless ad-hoc networks. The threshold constraint has a major effect on the routing decision. The $\lambda$ variation has a huge impact on the result; as there is a direct correlation between the $\lambda$ value and the number of lost packets, the higher $\lambda$ value the more packet loss, which result the system instability.

The topology control problem was discussed and formulated using integer linear programming. The result of the experiment demonstrated the presence of un-predictable traffics as there is an ability to prevent QoS requests. Additionally, the effects of threshold constraint on the sent packets through the network proves the higher the constraint number becomes the less possibility of losing packets and maximize the sensor Ad hoc WSN overall performance.

Finally, to keep the topology at its optimal level, the proposed algorithm for topology control should run iteratively to achieve both, the reduction of energy consumption and improving the lifetime of the whole network.

## 8. References

[1] Rorato, O., Bertoldo, S., Lucianaz, C., Allegretti, M., & Notarpietro, R. (2013). An Ad-Hoc Low Cost Wireless Sensor Network for Smart Gas Metering. Wireless Sensor Network, 05(03), 61-66. doi: 10.4236/ADHOC WSN.2013.53008

[2] J. Marchang, R. Douglas, B. Ghita, D. Lancaster and B. Sanders, "Dynamic Neighbour Aware Power-controlled MAC for Multi-hop Ad-hoc networks", Ad Hoc Networks, vol. 75-76, pp. 119-134, 2018. Available: 10.1016/j.adhoc.2018.04.003.

[3] R. Poovendran and L. Lazos, "A graph theoretic framework for preventing the wormhole attack in wireless ad hoc networks", Wireless Networks, vol. 13, no. 1, pp. 27-59, 2006. Available: 10.1007/s11276-006-3723-x.

[4] V. Hejlová and V. Voženílek, "Wireless Sensor Network Components for Air Pollution Monitoring in the Urban Environment: Criteria and Analysis for Their Selection", Wireless Sensor Network, vol. 05, no. 12, pp. 229-240, 2013. Available: 10.4236/ADHOC WSN.2013.512027.

[5] P. Kumar Singh and K. Lego, "Comparative Study of Radio Propagation and Mobility Models in Vehicular Adhoc Network", International Journal of Computer Applications, vol. 16, no. 8, pp. 37-42, 2011. Available: 10.5120/2031-2600.

[6] L. Rodrigues, C. Montez, G. Budke, F. Vasques and P. Portugal, "Estimating the Lifetime of Wireless Sensor Network Nodes through the Use of Embedded Analytical Battery Models", Journal of Sensor and Actuator Networks, vol. 6, no. 2, p. 8, 2017. Available: 10.3390/jsan6020008.

[7] D. Yadav, S. Chouhan, S. Vishvakarma and B. Raj, "Application Specific Microcontroller Design for Internet of Things Based Wireless Sensor Network", Sensor Letters, vol. 16, no. 5, pp. 374-385, 2018. Available: 10.1166/sl.2018.3965.

[8] Y. Gao, H. Wang, N. Weng and L. Vespa, "Enhancing Sensor Network Data Quality via Collaborated Circuit and Network Operations", Journal of Sensor and Actuator Networks, vol. 2, no. 2, pp. 196-212, 2013. Available: 10.3390/jsan2020196.

[9] D. Su, "The Research of Sensors' Interface on Wireless Sensor Network", Advanced Materials Research, vol. 926-930, pp. 2630-2633, 2014. Available: 10.4028/www.scientific.net/amr.926-930.2630.

[10] R. C and S. P, "Intelligent Traffic Monitoring Based on Wireless Sensor Network", International Journal of Trend in Scientific Research and Development, vol. -2, no. -5, pp. 759-761, 2018. Available: 10.31142/ijtsrd15943.

[11] L. Rodrigues, C. Montez, G. Budke, F. Vasques and P. Portugal, "Estimating the Lifetime of Wireless Sensor Network Nodes through the Use of Embedded Analytical Battery Models", Journal of Sensor and Actuator Networks, vol. 6, no. 2, p. 8, 2017. Available: 10.3390/jsan6020008.

[12] D. saritha and R. Prashanthi, "An RTOS Architecture for Industrial Wireless Sensor Network Stacks with Multi-Processor Support", International Journal of Engineering and Computer Science, 2015. Available: 10.18535/ijecs/v4i8.65.

[13] M. Keskin, İ. Altınel, N. Aras and C. Ersoy, "Wireless sensor network lifetime maximization by optimal sensor deployment, activity scheduling, data routing and sink mobility", Ad Hoc Networks, vol. 17, pp. 18-36, 2014. Available: 10.1016/j.adhoc.2014.01.003.

[14] D. J.Shah and H. A. Arolkar, "Single Point Interface for Data Analysis in Wireless Sensor Networks", International Journal of Computer Applications, vol. 47, no. 9, pp. 22-26, 2012. Available: 10.5120/7217-0022.

[15] S. Duan, "Design and Development of Detection Node in Wireless Sensor Network Based on Neural Network", Advanced Materials Research, vol. 1022, pp. 292-295, 2014. Available: 10.4028/www.scientific.net/amr.1022.292.

[16] X. Cheng, X. Huang, D. Li, W. Wu and D. Du, "A polynomial-time approximation scheme for the minimum-connected dominating set in ad hoc wireless networks", Networks, vol. 42, no. 4, pp. 202-208, 2003. Available: 10.1002/net.10097.

[17] GEORGOULAS, D. and BLOW, K. (2009). Wireless Sensor Network Management and Functionality: An Overview. Wireless Sensor Network, 01(04), pp.257-267.

[18] Peterson, L. and Davie, B. (1998). Computer Networks: A System Approach. IEEE Communications Magazine, 36(5), pp.54-54.

[19] Tiwari, A., Ballal, P. and Lewis, F. (2007). Energy-efficient wireless sensor network design and implementation for condition-based maintenance. ACM Transactions on Sensor Networks, 3(1), p.1-es.

[20] A. Manjeshwar, D.P. Agrawal, TEEN: a protocol for enhanced efficiency in wireless sensor networks, in: Proceedings of the 1st International Workshop on Parallel and Distributed Computing Issues in Wireless Networks and Mobile Computing, San Francisco, CA, April 2001.

[21] Y. Yu, D. Estrin, and R. Govindan, "Geographical and Energy-Aware Routing: A Recursive Data Dissemination Protocol for Wireless Sensor Networks," UCLA Comp. Sci. Dept. tech. rep., UCLA-CSD TR-010023, May 2001.

[22] Cui, S., Goldsmith, A., & Bahai, A. (2004). Energy-Efficiency of MIMO and Cooperative MIMO Techniques in Sensor Networks. IEEE Journal on Selected Areas in Communications, 22(6), 1089-1098. doi: 10.1109/jsac.2004.830916

[23] Kadayif, I.; Kandemir, M. Tuning In-Sensor Data Filtering to Reduce Energy Consumption in Wireless Sensor Networks. In Proceedings of Design, Automation and Test in Europe Conference and Exhibition, Paris, France, February 2004; pp. 1530-1539.

[24] Gupta D, Sampalli S. Cluster-head election using fuzzy logic for wireless sensor networks. In: IEEE conference on communication networks and services research (CNSR), Halifax, Novia Scotia, Canada; 2005. p. 255–60.

[25] María Gabriela Calle Torres, Energy Consumption In Wireless Sensor Networks Using GSP, University of Pittsburgh, Master of Science in Telecommunications, April, 2006.

[26] F. Shu, T. Sakurai, H. L. Vu and M. Zukerman, "WSN10-4: A Framework to Minimize Energy Consumption for Wireless Sensor Networks," IEEE Globecom 2006, San Francisco, CA, 2006, pp. 1-5. doi: 10.1109/GLOCOM.2006.949

[27] F. Shebli, I. Dayoub, A. O. M'foubat, A. Rivenq and J. M. Rouvaen, "Minimizing energy consumption within wireless sensors networks using optimal transmission range between nodes," 2007 IEEE International Conference on Signal Processing and Communications, Dubai, 2007, pp. 105-108. doi: 10.1109/ICSPC.2007.4728266

[28] G. Küçük and C. Başaran. Reducing energy consumption of wireless sensor networks through processor optimizations. Journal of Computers, 2(5):67–74, 2007

[29] T. Wang, W. Heinzelman and A. Seyedi, "Minimization of transceiver energy consumption in wireless sensor networks with AWGN channels," 2008 46th Annual Allerton Conference on Communication, Control, and Computing, Urbana-Champaign, IL, 2008, pp. 62-66. doi: 10.1109/ALLERTON.2008.4797536

[30] M. N. Halgamuge, M. Zukerman, K. Ramamohanarao, and H. L. Vu, "An Estimation of Sensor Energy Consumption," Progress In Electromagnetics Research B, Vol. 12, 259-295, 2009. doi:10.2528/PIERB08122303

[31] Y. Liang and W. Peng, "Minimizing energy consumptions in wireless sensor networks via two-modal transmission," ACM Comput. Commun. Rev., vol. 40, no. 1, pp. 12–18, Jan. 2010.

[32] Y. Wang, M. C. Vuran and S. Goddard, "Stochastic Analysis of Energy Consumption in Wireless Sensor Networks," 2010 7th Annual IEEE Communications Society Conference on Sensor, Mesh and Ad Hoc Communications and Networks (SECON), Boston, MA, 2010, pp. 1-9. doi: 10.1109/SECON.2010.5508259

[33] J. M. Molina, J. Haase, and C. Grimm, "Energy consumption estimation and profiling in wireless sensor networks," in Proc. 23th Int. Conf. Architecture of Comput. Syst. Workshop, ARCS'10, Feb. 2010, pp. 259–264.

[34] X. Wang, X. Wang and J. Zhao, "Impact of Mobility and Heterogeneity on Coverage and Energy Consumption in Wireless Sensor Networks," 2011 31st International Conference on Distributed Computing Systems, Minneapolis, MN, 2011, pp. 477-487. doi: 10.1109/ICDCS.2011.17

[35] K. Oikonomou and S. Aissa, "Dynamic sink assignment for efficient energy consumption in wireless sensor networks," 2012 IEEE Wireless Communications and Networking Conference (WCNC), Shanghai, 2012, pp. 1876-1881. doi: 10.1109/WCNC.2012.6214091

[36] S. G. Jia et al., "An Efficient Sleeping Scheduling for Save Energy Consumption in Wireless Sensor Networks", Advanced Materials Research, Vols. 756-759, pp. 2288-2293, 2013

[37] Tony Ducrocq, Michaël Hauspie, and Nathalie Mitton, "Balancing Energy Consumption in Clustered Wireless Sensor Networks," ISRN Sensor Networks, vol. 2013, Article ID 314732, 14 pages, 2013. https://doi.org/10.1155/2013/314732.

[38] Nizar HA, Tarik ZI, Rassim NI (2013) Optimization of Energy Consumption in Wireless Sensor Networks based on Nature-Inspired Algorithms. IJCA 77: 32-39.

[39] Saraswat J (2013) "Effect of duty cycle on energy consumption in Wireless Sensor Networks". Int J Computer Netw Commun5(1):125–140

[40] Loshakov, VA & Al-Janabi, Hussam & Hussein, Yahya. (2013). Improving Energy Consumption in Wireless Sensor Networks by LTE with MIMO.

[41] H. Cotuk, B. Tavli, K. Bicakci and M. B. Akgun, "The impact of bandwidth constraints on the energy consumption of Wireless Sensor Networks," 2014 IEEE Wireless Communications and Networking Conference (WCNC), Istanbul, 2014, pp. 2787-2792. doi: 10.1109/WCNC.2014.6952870

[42] Nandi, Madhusmita & Roy, Jibendu. (2014). Optimization of Energy Consumption in Wireless Sensor Networks Using Particle Swarm Optimization. International Journal of Computer Application (IJCA). 91. 45-50. 10.5120/15961-5418.

[43] Catarinucci L, Colella R, Del Fiore D, et al. A cross-layer approach to minimize the energy consumption in wireless sensor networks. Int J Distrib Sens N. Epub ahead of print 1 January 2014. DOI: 10.1155/2014/268284.

[44] Nejah, Nasri & Kachouri, Abdennaceur. (2014). Impact of Topology on Energy Consumption in Wireless Sensor Networks. Journal of Machine-to-Machine Communications. 1. 145-160.

[45] Hou R., Zheng M., Chang Y., ET AL.: 'Analysis of battery energy consumption in wireless sensor networks considering path correlation', Sens. Lett., 2015, 13, (3), pp. 240–244

[46] Yazdanpanah, S., Varzi, Y., Haronabadi, A & Mirabedini, S. (2014). An improved energy consumption method for wireless sensor networks.Management Science Letters, 4(6), 1123-1132.

[47] A. Girgiri, M. A. Baba, L. G. Ali and M. Adamu, "Minimising energy consumption in wireless sensor network by enhancement of leach protocol," 2015 12th International Joint Conference on e-Business and Telecommunications (ICETE), Colmar, 2015, pp. 53-61.

[48] M. Lounis, A. Bounceur, A. Laga and B. Pottier, "GPU-based parallel computing of energy consumption in wireless sensor networks," 2015

European Conference on Networks and Communications (EuCNC), Paris, 2015, pp. 290-295. doi: 10.1109/EuCNC.2015.7194086

[49] S. Yaessad, L. Bouallouche, and D. Aissani, "A Cross-Layer Routing Protocol for Balancing Energy Consumption in Wireless Sensor Networks" Wireless Pers. Commun., Springer, 2014.

[50] B. Amutha, B. Ghanta, K. Nanamaran, M. Balasubramanian, ECOSENSE: an energy consumption protocol for wireless sensor networks. Procedia Comput. Sci. 57, 1160–1170 (2015)

[51] Mohammad Mortazavi, Seyyed & Malekzadeh, Mina & Askari, Reza. (2016). Reducing Energy Consumption in Wireless Sensor Networks by CICA Algorithm. 3021-3031.

[52] Viktor Toldov, Román Igual-Pérez, Rahul Vyas, Alexandre Boé, Laurent Clavier, et al. Experimental Evaluation of Interference Impact on the Energy Consumption in Wireless Sensor Networks. 17th IEEE International Symposium on a World of Wireless, Mobile and Multimedia Networks (WoWMoM), Jun 2016, Coimbra, Portugal.

[53] H. Oudani et al., "Minimize energy consumption in wireless sensor network using hierarchical protocols," 2017 International Conference on Engineering & MIS (ICEMIS), Monastir, 2017, pp. 1-9. doi: 10.1109/ICEMIS.2017.8272969

[54] Okopa, Michael et al. "Modeling Delay And Energy Consumption For Wireless Sensor Networks With High Coefficient Of Variability". Australasian Journal Of Computer Science, vol 4, no. 1, 2017, pp. 17-31. Science Alert, doi:10.3923/aujcs.2017.17.31.

[55] Alhumud, Haythem, and Mohammed Zohdy. "Managing Energy Consumption Of Wireless Sensors Networks In Multiple Greenhouses". Wireless Engineering and Technology, vol 09, no. 02, 2018, pp. 11-19. Scientific Research Publishing, Inc, doi:10.4236/wet.2018.92002.

[56] R. Alhussaini, A. K. Idrees, M. A. Salman, "Data transmission protocol for reducing the energy consumption in wireless sensor networks", International Conference on New Trends in Information and Communications Technology Applications, pp. 35-49, 2018.

[57] Cundeva-Blajer, Marija and Mare Srbinovska. "Mathematical Tools for Optimization of Energy Consumption in Wireless Sensor Networks." (2018).

[58] K S Ananda Kumar, Balakrishna R "Implementation of Itree-MAC Protocol for Effective Energy Consumption in Wireless Sensor Networks", International Journal of Scientific Research in Computer Science, Engineering and Information Technology (IJSRCSEIT), pp. 113-117, 2018

[59] Tabaa, M. (2016). A Novel Transceiver Architecture Based on Wavelet Packet Modulation for UWB-IR ADHOC WSN Applications. Wireless Sensor Network, 08(09), pp.191-209.

[60] Chen, X., Makki, K., Yen, K. and Pissinou, N. (2009). Sensor network security: a survey. IEEE Communications Surveys & Tutorials, 11(2), pp.52-73.

[61] Aggelou, G. (2004). On the Performance Analysis of the Minimum-Blocking and Bandwidth-Reallocation Channel-Assignment (MBCA/BRCA) Methods for Quality-of-Service Routing Support in Mobile Multimedia Ad Hoc Networks. IEEE Transactions on Vehicular Technology, 53(3), pp.770-782.

[62] M. Yarvis and M. Zorzi, "Ad hoc networks: Special issue on energy efficient design in wireless ad hoc and sensor networks", Ad Hoc Networks, vol. 6, no. 8, pp. 1183-1184, 2008. Available: 10.1016/j.adhoc.2007.11.005.

[63] F. Alassery, W. Ahmed, M. Sarraf and V. Lawrence, "A Novel Low-Complexity Low-Latency Power Efficient Collision Detection Algorithm for Wireless Sensor Networks", Wireless Sensor Network, vol. 07, no. 06, pp. 43-75, 2015. Available: 10.4236/ADHOC WSN.2015.76006.

[64] N. Zhang and A. Anpalagan, "Comparative Review of QoS-Aware On-Demand Routing in Ad Hoc Wireless Networks", Wireless Sensor Network, vol. 02, no. 04, pp. 274-284, 2010. Available: 10.4236/ADHOC WSN.2010.24038.

[65] D. Li, X. Jia and H. Du, "QoS Topology Control for Nonhomogenous Ad Hoc Wireless Networks", EURASIP Journal on Wireless Communications and Networking, vol. 2006, pp. 1-10, 2006. Available: 10.1155/wcn/2006/82417.